\documentclass{eas}

\usepackage{amssymb}
\usepackage{graphicx}
\usepackage{mathptmx}
\usepackage{url}

\newcommand\arcdeg{\mbox{$^\circ$}}
\newcommand\arcmin{\mbox{$^\prime$}}
\newcommand\arcsec{\mbox{$^{\prime\prime}$}}

\runningtitle{Solar Physics and the Solar-Stellar Connection at Dome C}

\TitreGlobal{2$^{\rm nd}$ ARENA Conference on ``The Astrophysical Science
    Cases at Dome C''}
\Editors{H.\ Zinnecker, H.\ Rauer, and N.\ Epchtein (eds.)}
\idline{EAS Publications Series, Vol.~??, 2008}
\doi{10.1051/eas:2008}
\FirstPage{1}

\begin{document}

\title{Solar Physics and the Solar-Stellar Connection at Dome C}
\author{Carsten Denker$^{1}$}
\author{Klaus G.\ Strassmeier}
\address{Astrophysikalisches Institut Potsdam,
    An der Sternwarte 16,
    D-14482 Potsdam,
    Germany}

%==============================================================================
%    ABSTRACT
%==============================================================================

\begin{abstract}
Solar magnetic fields evolve on many time-scales, \textit{e.g.}, the
generation, migration, and dissipation of magnetic flux during the 22-year
magnetic cycle of the Sun. Active regions develop and decay over periods of
weeks. The build-up of magnetic shear in active regions can occur within less
than a day. At the shortest time-scales, the magnetic field topology can change
rapidly within a few minutes as the result of eruptive events such as flares,
filament eruptions, and coronal mass ejections. The unique daytime seeing
characteristics at \textit{Dome~C}, \textit{i.e.}, continuous periods of very
good to excellent seeing during almost the entire Antarctic summer, allow us to
address many of the top science cases related to the evolution of solar
magnetic fields. We introduce the \textit{Advanced Solar Photometric Imager and
Radiation Experiment} and present the science cases for synoptic solar
observations at \textit{Dome~C}. Furthermore, common science cases concerning
the solar-stellar connection are discussed in the context of the proposed
\textit{International Concordia Explorer Telescope}.
\end{abstract}
\maketitle

%==============================================================================
%    INTRODUCTION
%==============================================================================

\section{Introduction}

Ground-based solar physics is currently seeing vigorous efforts to upgrade
existing facilities and to develop the next generation of telescopes and
instruments. These efforts are not restricted to telescopes with large
apertures (1.5 to 4~m) for high-resolution observations of solar fine-structure
but also include synoptic instruments and special purpose telescopes such as
coronagraphs.

Several major solar telescope are currently being build \textit{e.g.}, the
\textit{New Solar Telescope} (NST) at Big Bear Solar Observatory (BBSO) in
California (Denker \etal\ \cite{Denker2006}) and the German \textit{GREGOR}
project at Observatorio del Teide on Tenerife, Spain (Volkmer \etal\
\cite{Volkmer2006}). The \textit{Advanced Technology Solar Telescope} (ATST,
Wagner \etal\ \cite{Wagner2006}) under the stewardship of the U.S.\ National
Solar Observatory (NSO) is in the final stages of the design and development
phase awaiting funding approval for construction at Mees Solar Observatory
(MSO) on Maui, Hawai'i. The current suite of synoptic telescopes includes the
six \textit{Global Oscillation Network Group} (GONG, Harvey \etal\
\cite{Harvey1996}) observing stations to investigate the internal structure and
dynamics of the Sun using helioseismology, the \textit{Synoptic Optical
Long-term Investigations of the Sun} (SOLIS, Keller, Harvey and Giampapa
\cite{Keller2003}) facility at Kitt Peak National Observatory (KPNO) in Arizona
to study the solar activity cycle, the two \textit{Precision Solar Photometric
Telescopes} (PSPTs, Coulter and Kuhn \cite{Coulter1994}) at Mauna Loa Solar
Observatory (MLSO) on the Big Island of Hawai'i and Osservatorio Astronomico di
Roma (OAR) in Italy to monitor solar irradiance changes, and the
\textit{Optical Solar Patrol Network} (OSPAN), formerly known as the
\textit{Improved Solar Observing Optical Network} (ISOON, Neidig \etal\
\cite{Neidig1998}), to monitor solar eruptive events. Finally, a meter-class
coronagraph has been proposed by the High-Altitude Observatory (HAO) of the
National Center for Atmospheric Research (NCAR) in Boulder, Colorado. The
\textit{Coronal Solar Magnetism Observatory} (COSMO, Tomczyk, Lin, and
Zurbuchen \cite{Tomczyk2007}) would replace existing MLSO facilities.

Astronomical seeing conditions at \textit{Dome~C} are exceptional (Lawrence
\etal\ \cite{Lawrence2004}) and very promising for solar research. However, any
new solar project has to establish a compelling science case, which exploits
the unique seeing characteristics of the Antarctic plateau and which is
competitive in this highly active research environment.

%==============================================================================
%    DAYTIME SEEING CHARACTERISTICS AT DOME C
%==============================================================================

\section{Daytime Seeing Characteristics at Dome~C}

The French/Italian \textit{Concordia} station at \textit{Dome~C} is located on
the Antarctic plateau (75\arcdeg 6\arcmin\ South and 123\arcdeg 21\arcmin\
East) at an elevation of 3250~m. Aristidi \etal\ (\cite{Aristidi2005}) carried
out a rigorous study of the daytime seeing characteristics at this site. The
data were obtained with a differential image motion monitor (DIMM,
\textit{e.g.}, Sarazin and Roddier 1990) during two 3-months campaigns during
the Antarctic summers of 2003--2004 and 2004--2005. They measured a median
seeing of 0.54\arcsec\ and a medium isoplanatic angle of 6.8\arcsec, which
indicate excellent seeing conditions. The seeing was typically even better for
several hours during the afternoon ($\approx 0.4$\arcsec\ corresponding to a
Fried-parameter $r_0 \simeq 25.0$~cm at 500~nm). However, the Antarctic daytime
seeing measurements were based an observations of the bright star Canopus and
the aforementioned results have been zenith angle corrected, which is not
suitable evaluating the potential for solar observations. Therefore, the seeing
statistics have to be reevaluated for solar observations and interpreted in the
light of a specific science case.

Seeing with a Fried-parameter better than $r_0 \gtrsim 7.0$~cm marks the
threshold at which wavefront sensors can lock on granulation as a target and
solar adaptive optics (AO) systems can operate (Rimmele \cite{Rimmele2000}).
The excellent seeing regime starts with Fried-parameters $r_0 \gtrsim 12.0$~cm.
This enables high spectral resolution observations, which require long exposure
times (in excess of the daytime coherence time $t_{\rm exp} \gtrsim \tau_0
\approx 40.0$~ms). Good time coverage is required to follow the evolution of
three-dimensional flow and magnetic fields in active regions. An interesting
statistical property is the number of hour blocks when the seeing continuously
exceeds the thresholds of $r_0 = 7.0$ and 12.0~cm, respectively. Anecdotal
evidence is presented in Aristidi \etal\ (\cite{Aristidi2005}) who report
exceptional seeing as low as 0.1\arcsec\ during a 10-hour continuous period of
seeing below 0.6\arcsec.

\textit{Concordia} station provides a range of seeing characteristics, which
are attractive for solar observations: The large clear time fraction (CTF) of
at least 75\% during the three months of Antarctic summer promises about 1,600
hours of observations. This is about 60\% of the total observing time of
low/mid-latitude solar observatories, which observe the whole year. However, it
might well be, if the hour block statistics for the $r_0 = 7.0$ and 12.0~cm
thresholds are carried out, that \textit{Concordia} station outperforms other
solar sites. The final report of the ATST site survey working group could serve
as a benchmark (Hill \etal\ \cite{Hill2004}). During the Antarctic summer, the
Sun ``oscillates'' above the horizon, providing long continuous time sequences,
without interruptions due to the day-night cycle. This is a unique property,
which can otherwise only be achieved from space.

The excellent seeing conditions (large Fried-parameter, large isoplanatic
angle, and long atmospheric coherence time) are advantageous for high spatial
resolution observations. However, synoptic full-disk observations would also
benefit, since the normalized Fried-parameter $\alpha = r_0 / D$, where $D
\approx 25.0$~cm is the telescope diameter, would be close to unity.
Furthermore, low-order wavefront correction facilitated by ground-layer
adaptive optics (GLAO, Tokovinin \cite{Tokovinin2004}) might become feasible.
Aristidi \etal\ (\cite{Aristidi2005}) find that almost half of the ground
turbulence is concentrated into the first 5.0~m above the surface. The
performance of a GLAO system is not a strong function of the field of view
(FOV). Since low-altitude aberrations are isoplanatic (Tokovinin
\cite{Tokovinin2004}), only a thin layer has to be corrected at
\textit{Dome~C}. Thus, a large field can be compensated. A wide-field GLAO
system with a $5\arcmin$ FOV has been proposed for the \textit{Gemini}
telescopes (Andersen \etal\ \cite{Andersen2006}).

Finally, the extremely cold and dry air, very low infrared sky emission, and
low aerosol and dust content will offer some of the best coronal skies
(Lawrence \etal\ \cite{Lawrence2004}, Kenyon and Storey \cite{Kenyon2006}).
Efforts are currently underway, to quantify the site characteristics for
coronal observations.

%==============================================================================
%    ADVANCED SOLAR PHOTOMETRIC IMAGER AND RADIATION EXPERIMENT
%==============================================================================

\section{Advanced Solar Photometric Imager and Radiation Experiment}

\begin{figure}[t]
\centerline{\includegraphics[width=0.9\textwidth]{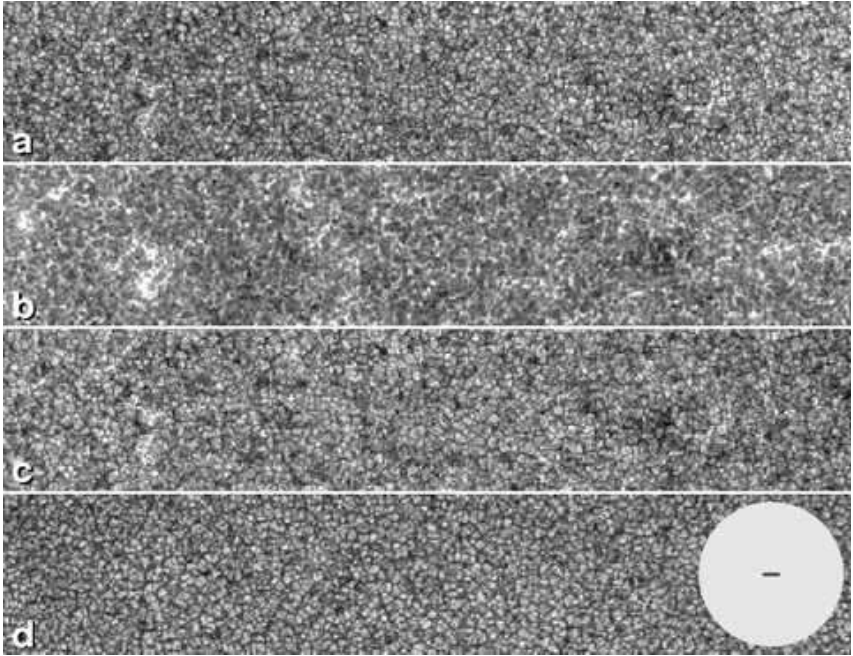}}
\caption{Quiet Sun observations obtained with the Broadband Filter Imager
    (BFI) of the \textit{Solar Optical Telescope} (SOT) on board \textit
    {Hinode}(Kosugi \etal\ \cite{Kosugi2007}) on 2007 September~5. The
    filtergrams were taken quasi-simultaneously in \textbf{a} the CN bandhead
    (388.4~nm), \textbf{b} the strong chromospheric absorption line
    Ca\,{\sc ii}\,H (396.9~nm), \textbf{c} the G-band (430.5~nm), and
    \textbf{d} the blue continuum (450.5~nm), respectively. The FOV is
    $220\arcsec \times 42\arcsec$. About nine of these image stripes have to be
    assembled to cover the entire solar equator. The insert in the corner
    provides a comparison of the solar disk with the high-resolution FOV.}
\label{FIG01}
\end{figure}

The envisioned \textit{Advanced Solar Photometric Imager and Radiation
Experiment} (ASPIRE) is an innovative photometric full-disk telescope to
explore the solar photosphere and chromosphere in the blue spectral region.
ASPIRE could share the same mount and use the infrastructures of the
\textit{International Concordia Explorer Telescope} (ICE-T), thus, leveraging
the investments in ICE-T by adding a daytime science component. The ASPIRE
concept is well adapted to the seeing conditions at \textit{Dome~C} and relates
to a variety of promising science cases. The blue end of the visible spectrum
contains a variety of interesting spectral features (see Fig.~\ref{FIG01}).
Smart coatings on the entrance window will restrict the transmitted sunlight to
380--460~nm. Thus, only 10\% of the solar radiant energy will enter the
instrument, which is less than 10~W. Thermal loads on optical elements are not
critical but an evacuated or helium filled telescope tube has to be considered
to avoid instrument seeing. Relatively wide ($\approx 1.0$~nm) interference
filters can be used to isolate these spectral regions, thus, enabling
photometric tomography (height dependence) of small-scale magnetic features.

The key idea of ASPIRE is to combine synoptic observations with high spatial
resolution. In addition, ASPIRE data will be obtained with high temporal
resolution (one image per spectral region every 30~s) as well as a low
scattered light level and very good photometric accuracy (better than 0.1\%).
As part of ICE-T, large-format detectors with 10k $\times$ 10k pixels will
become available for daytime observations. Full-disk filtergrams would have an
image scale of $0.2\arcsec$ pixel$^{-1}$ with such detectors. Considering the
excellent seeing conditions at \textit{Dome~C}, telescopes with small apertures
($\approx 25.0$~cm) could obtain images with a spatial resolution better than
0.5\arcsec\ during substantial periods of time. Assuming an exposure time of
10~ms, a bandpass of 1.0~nm and a total wavelength-dependent system efficiency
of 2.5--3.3\%, the photon statistics ($n > 200,000$) would be well adapted to
the full well capacity of the CCD and still leave some latitude in filter
specifications and observing procedures. Short exposure times are necessary to
allow for post-facto image restoration in addition to real-time GLAO
correction. A small number of filtergrams (3--10) is sufficient for multi-frame
blind deconvolution (MFBD) techniques (\textit{e.g.}, van Noort, Rouppe van der
Voort, and L\"ofdahl \cite{vanNoort2005}).

\begin{figure}[t]
\centerline{\includegraphics[width=0.95\textwidth]{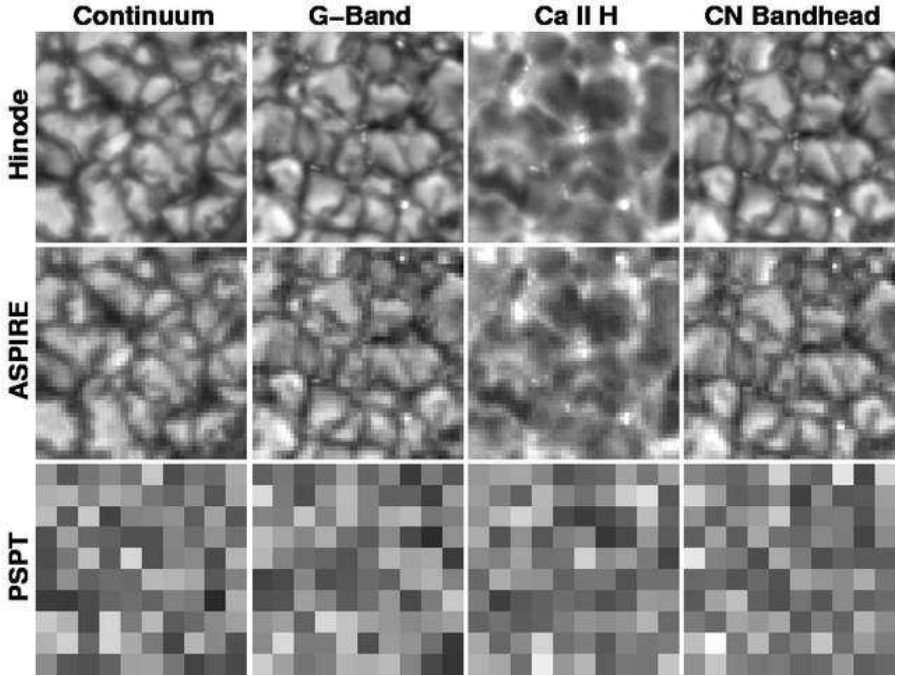}}
\caption{Comparison of the spatial resolving power between \textit{Hinode} SOT,
    ASPIRE, and PSPT for different spectral regions. The FOV is $10\arcsec
    \times 10\arcsec$.}
\label{FIG02}
\end{figure}

In Fig.~\ref{FIG02}, we show a comparison of the spatial resolution, which can
be achieved with \textit{Hinode} SOT, ASPIRE, and the \textit{Precision Solar
Photometric Telescope} (PSPT, Coulter and Kuhn \cite{Coulter1994}). The image
scales of these instruments are 0.054, 0.2, and 1.0\arcsec\ pixel$^{-1}$,
respectively. We degraded the \textit{Hinode} SOT data to match the image scale
of the other instruments. The resolving power of ASPIRE is sufficient to
resolve solar granulation and even small-scale bright points are still
detectable. Therefore, photospheric and chromospheric optical flows can be
measured, \textit{i.e.}, horizontal proper motions, across the entire disk.
Furthermore, feature identification and pattern recognition techniques can be
applied to the filtergrams. These capabilities are currently not available from
any other synoptic telescope.

High spatial and temporal resolution come at a price. A single 16-bit image is
about 200~MB, which results in a data rate of 1.6~GB min$^{-1}$. About 250~TB
data would be collected over one Antarctic summer. This huge amount of data
might be hard to handle. We intend to save one filtergram per spectral region
with full spatial resolution every 30~min. This would result in a total of
about 2~TB of data (with lossless compression). High cadence observations are
primarily required for local correlation and feature tracking to measure the
optical flows (\textit{e.g.}, November and Simon \cite{November1988}, Roudier
\etal\ \cite{Roudier1999}). In this case, photometric accuracy can be
sacrificed and lossy compression algorithms can be used. The data amount with
full spatial and temporal resolution would be in this case only 25~TB. In
summary, depending on the scientific objective observing modes can be selected,
which fit the data storage and analysis limitations.

%==============================================================================
%    SCIENCE CASES FOR SYNOPTIC SOLAR OBSERVATIONS
%==============================================================================

\section{Science Cases for Synoptic Solar Observations}

\subsection{Differential Rotation, Meridional Flows, and Large-Scale Flow
    Patterns}

Horizontal proper motions can be detected as optical flows using local
correlation or feature tracking algorithms. The final resolution of the
velocity maps will match \textit{Helioseismic and Magnetic Imager} (HMI) data,
which will become available in August 2008 with the launch of the \textit{Solar
Dynamics Observatory} (SDO, Kosovichev \etal\ \cite{Kosovichev2007}). The flow
field contains contributions from solar differential rotation, meridional
flows, and localized flow patterns due to active regions evolution and flux
emergence. Seeing will introduce a noise term so that between 50 and 100
filtergrams are needed for one flow map. Photometric tomography then provides
the height dependence of the flow fields. The transition from differential
surface rotation to the more rigid rotation of the corona is one topic, which
can be addressed with ASPIRE. Coronal hole regions are of particular interest
and UV/EUV filtergrams obtained with the SDO's \textit{Atmospheric Imaging
Assembly} (AIA) can provide the coronal context. Since the velocity signal
(300--500~m~s$^{-1}$) of supergranulation is mostly horizontal, ASPIRE would be
ideal to study large scale convective patterns and giant cells. These
convective motions can than be related to the Ca\,{\sc ii}\,K network and other
features associated with the Sun's ubiquitous magnetic field.

The solar surface is highly dynamic, and magnetic fields on various spatial
scales, from active regions and sunspots to network fields and faculae, are
intricately intertwined with plasma motions. Photospheric flows can contribute
to the build-up of magnetic shear in active regions, which will store energy in
the complex three-dimensional magnetic field topology above active regions.
This energy can be violently released in flares, filament and prominence
eruptions, and coronal mass ejections (CMEs). The importance of plasma flows
associated with solar eruptive events was recently demonstrated by Yang \etal\
(\cite{Yang2004}). ASPIRE's global coverage will help to monitor flow patterns
in active regions, which might become source regions for solar eruptive events.

\subsection{Solar-Terrestrial Relations}

The total radiative energy arriving at Earth is known as the solar ``constant''
$S = 1366 \pm 3$~W~m$^{-2}$. On the other hand, solar activity has a pronounced
11-year cycle which manifests itself in the radio flux, the X-ray background,
the UV/EUV and total irradiance and the frequency of various solar features
such as sunspots, plages, filaments, prominences, flares, CMEs, and coronal
holes. The common source for solar variability is the Sun's magnetic field.
Analyzing the magnetic field data reveals that the magnetic cycle is actually
22 years, switching from a North-South to a South-North configuration and back.
This apparent dichotomy between constancy and variability makes the radiative
coupling of the Sun-Earth system an intriguing field of research. Especially,
since solar irradiance variability has been related to global climate change on
Earth and the discussion of its importance compared to anthropogenic effects.
The UV and EUV show the strongest irradiance variations. Faculae and plages are
particularly strong emitters in this specific wavelength region. Furthermore,
UV/EUV radiation plays an important role in the energy balance and chemistry of
Earth's upper atmospheric layers, \textit{e.g.}, the dissociation of ozone by
UV photons in the range of 200 to 300~nm. Ca\,{\sc ii}\,K filtergrams provide
an important proxy for the UV/EUV variability. A detailed review of solar
irradiance variability and solar–terrestrial effects was given by Lean
(\cite{Lean1997}) and more recently by Fr\"ohlich and Lean
(\cite{Froehlich2004}).

Feature identification and pattern recognition can be used to discriminate
between plages, faculae, network and sunspots and thus measure their
contribution to irradiance variability. Brightness temperatures are accessible
with multi-color photometry and the flux tube geometry can be mapped in height
by studying the center-to-limb variation of faculae, pores, plages, and
filigree. This structural analysis will be complemented by high resolution HMI
vector magnetograms. ASPIRE is envisioned as a pathfinder for solar
observations at \textit{Dome~C}. However, its contributions towards solar cycle
variations are significant, which could warrant an extension of the project and
justify funding to cover an entire solar cycle.

\subsection{Solar-Stellar Connection}

ICE-T is a twin 60-cm aperture telescope for Sloan $g$ and $i$ photometry
(Strassmeier \etal\ \cite{Strassmeier2007,Strassmeier2008}). Precision
wide-field photometry is the common denominator of ICE-T and ASPIRE enabling
synergies in instrumentation, IT infrastructure and data analysis. Beyond the
instrumentation aspects, common ground can be found for a variety of science
cases such as (stellar) differential rotation, magnetic activity of late-type
stars, and dynamo activity. ASPIRE is in this respect a Sun-as-a-Star telescope
but with unprecedented spatial resolution.

Themes of the solar-stellar connection (Cayrel de Strobel
\cite{CayreldeStrobel1996}) include magnetic cycles of activity and their
evolution, atmospheric structure(s), heating processes, abundances, and the
presence of planets around cool stars other than the Sun (see \textit{e.g.},
Dupree \cite{Dupree2003}). \textit{ASPIRE} will measures globally horizontal
proper motions, thus, providing information on (1) differential rotation, which
converts poloidal to toroidal fields, (2) meridional flows at the solar
surface, which transport flux towards poles, and (3) convective surface
patterns, which lead to diffusion of the magnetic field. More and more details
of these solar dynamo signatures are now being extracted from and discovered in
stellar observations as well (see \textit{e.g.}, Strassmeier
\cite{Strassmeier2005}). Combining highly resolved ``snapshots'' of solar
activity with observations of solar-type stars at various evolutionary stages
demands models, which integrate these diverse data sets to answer a variety of
fundamental questions: How does the dynamo evolve over the life-time of a star
and how is it related to spin down and angular momentum transport by stellar
winds? How often do Maunder minimum events occur? What is the relationship
between differential rotation and rotation rate? Do counterparts to solar
activity nests and active longitudes exists and how do they affect the
structure of stellar coronae?

Finally, the emerging field of \textit{space climate} (Nandy and Martens
\cite{Nandy2007}) embraces the topics of solar-terrestrial relations and
solar-stellar connection. Modulation of the magnetic, radiative and particle
environment in the heliosphere is of great importance for star-planet
interactions in the habitable zone over stellar evolution time scales. Magnetic
activity changes as the dynamo evolves during the stars lifespan, thus, forcing
planetary systems -- which respond. The details of such an interaction of
course pique the human curiosity, since they are closely tied to the question
of the origin of life on Earth and possibly on other planets.

%==============================================================================
%    BIBLIOGRAPHY
%==============================================================================

\end{document}